\documentclass{IEEEtran}
\ifCLASSINFOpdf
\else
\fi
\usepackage{multicol}
\usepackage{lipsum}
\usepackage{xcolor}
\usepackage{multirow}
\usepackage{enumitem}
\usepackage{geometry}
\usepackage{pdfpages}
\usepackage{amsfonts}
\usepackage{amssymb}
\usepackage{amsmath}
\usepackage{graphicx}
\usepackage{float} 
\usepackage{subfigure}
\usepackage{lipsum}
\usepackage{multicol}
\usepackage{booktabs}

\makeatletter
\def\footnoterule{\kern-3\p@
	\hrule \@width 0.49\textwidth \kern 2.6\p@}
\makeatother

\usepackage{caption}
\newcommand{\subparagraph}{}
\usepackage{titlesec}
\titlespacing{\section}
{0pt}{5pt}{5pt}
\titlespacing{\subsection}
{0pt}{5pt}{5pt}

\begin{document}

% 不使用\maketitle命令生成标题，直接在文内按照Word文档的格式要求生成标题。
% 此处输入标题，fontsize{14}{14}第一个参数为字符大小为14号
% 第二个参数为字符间距，也为14.
\begin{flushleft}
	\fontsize{14}{15} \fontfamily{phv}\selectfont
	 \textbf{A Fast Method for Steady-State Memristor Crossbar Array Circuit Simulation
	}
\end{flushleft}

% 此处输入作者及所属机构
\begin{flushleft}	
	{\fontsize{10}{10} \fontfamily{phv}\selectfont 
		Rui Xie\textsuperscript{1}, Mingyang Song\textsuperscript{1}, Junzhuo Zhou\textsuperscript{1}, Jie Mei\textsuperscript{1}, Quan Chen{$^*$}\textsuperscript{1} (Corresponding Author) \\
		\textsuperscript{1}School of Microelectronics, Southern University of Science and Technology\\
		
% 		\textsuperscript{2}Name of 2\textsuperscript{nd} organization/affiliation (keep short, align left)
	}
% % 此处输入funding机构
% 	\blfootnote{\textcolor[rgb]{1,0,0}{Identify applicable funding agency here. If none, delete this text box}}

\end{flushleft}
	\noindent
	\begin{abstract}
	    In this work we propose an effective preconditioning technique to accelerate the steady-state simulation of large-scale memristor crossbar arrays (MCAs). We exploit the structural regularity of MCAs to develop a specially-crafted preconditioner that can be efficiently evaluated utilizing tensor products and block matrix inversion. Numerical experiments demonstrate the efficacy of the proposed technique compared to mainstream preconditioners. 
	\end{abstract}
	
	\noindent

\section{Introduction}

MCAs (Memristor Crossbar Arrays)~\cite{Chua:71} has gained substantial attention recent years because of its potential application in high-performance AI hardware and neuromorphic computing~\cite{Zhang:20}, calling for efficient circuit simulation tools. However, efficient simulation of MCAs has become increasingly challenging. The expected size of MCA is growing rapidly to accommodate the millions of weights involved in state-of-the-art neural networks~\cite{Truong:16}. Furthermore, a large amount of simulations are needed for statistical characterization or if the training/inference procedures are to be studied at circuit simulation level.  

Existing steady-state simulation of MCA circuits is often done by SPICE, in which a sparse linear system resulted from the modified nodal analysis (MNA) must be solved in each Newton iteration. The matrix size can be huge, e.g., a $1024 \times 1024$ MCA leads to a matrix size $> 10^6$, resulting in severe bottlenecks in time and memory consumption if direct solvers are used. Iterative solvers can improve the scalability, but existing general-purpose preconditioners~\cite{Ferronato:12} are often not adequately efficient for large-scale MCA circuits.     
In this work, we leverage the special topology of MCAs to develop an efficient preconditioning technique to accelerate the steady-state simulation of MCAs. Specifically, the preconditioner has the following features:
\begin{enumerate}
\item It takes advantages of the topological regularity of MCAs to generate special block structures;
\item Its inverse and application to vectors can be efficiently evaluated by Kronecker product and block matrix inversion formula.
\end{enumerate}

\section{Background}

% \subsection{MCA Circuit Structure}

\begin{figure}[h]
\centering  %图片全局居中
\subfigure[The middle layer of memristors.]{
\label{fig:div_mem}
\includegraphics[width=0.15\textwidth]{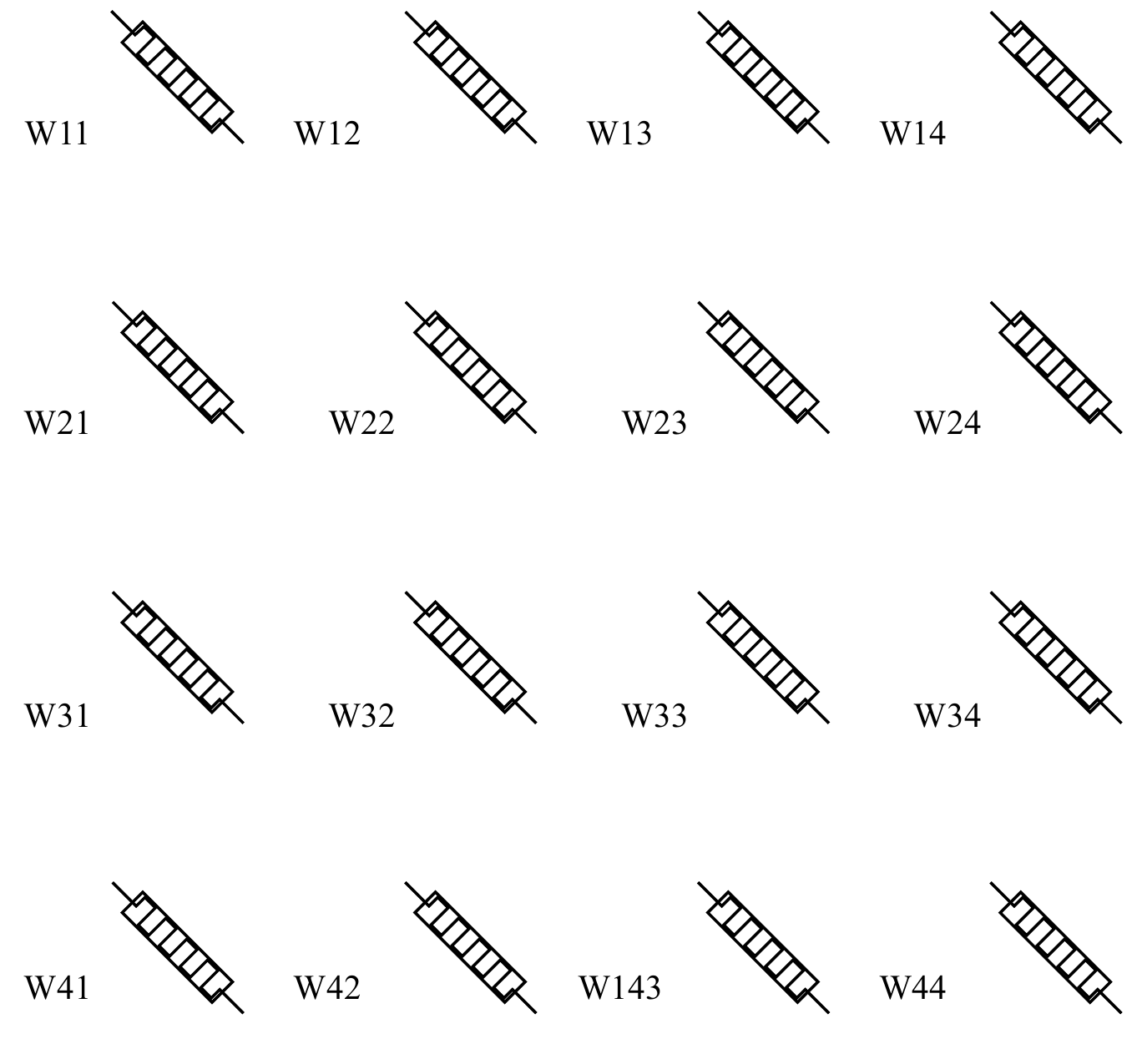}}
\subfigure[The top metal layer with conductance represented as $G_t$.]{
\label{fig:div_col}
\includegraphics[width=0.15\textwidth]{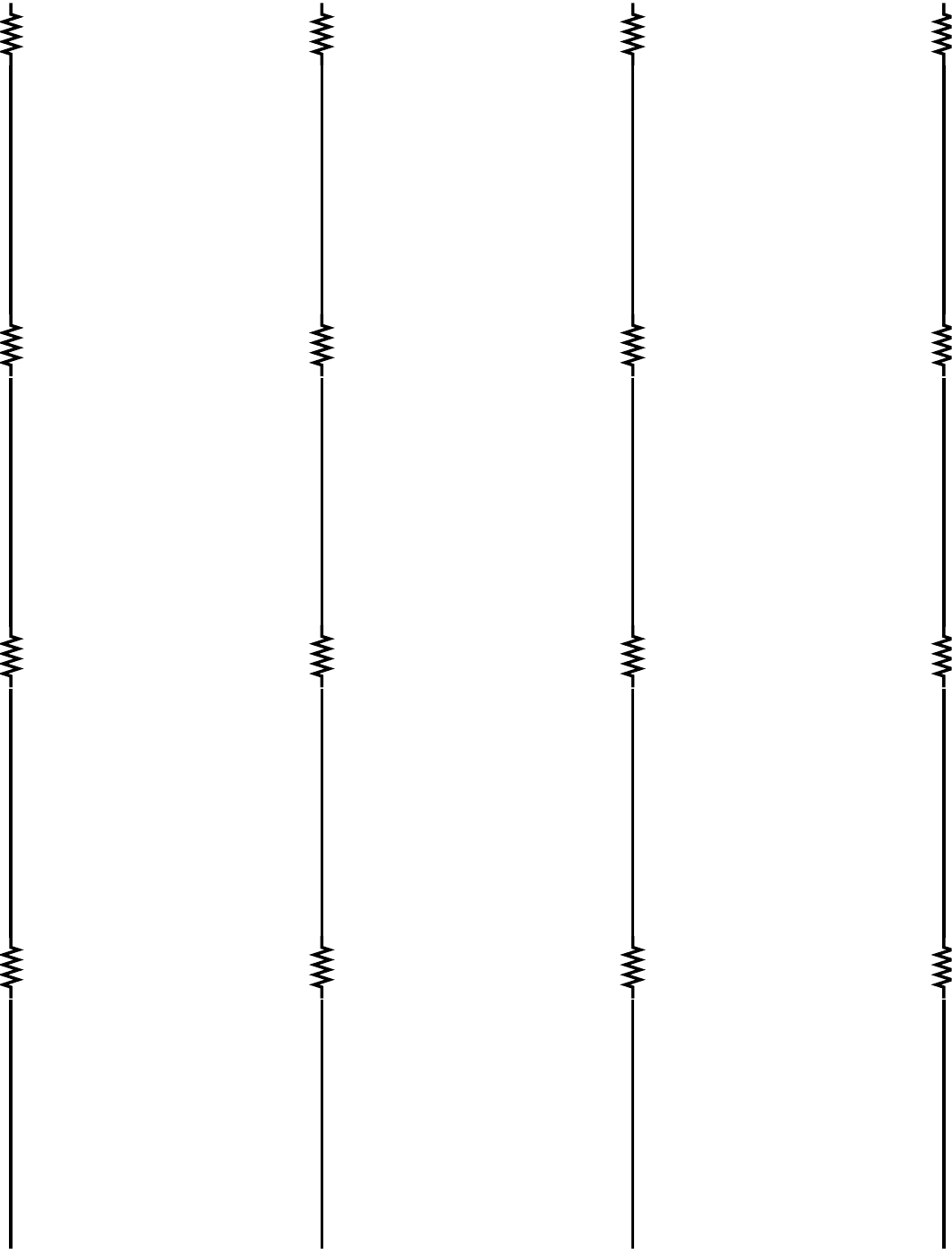}}
\subfigure[The bottom metal layer with conductance represented as $G_b$.]{
\label{fig:div_row}
\includegraphics[width=0.15\textwidth]{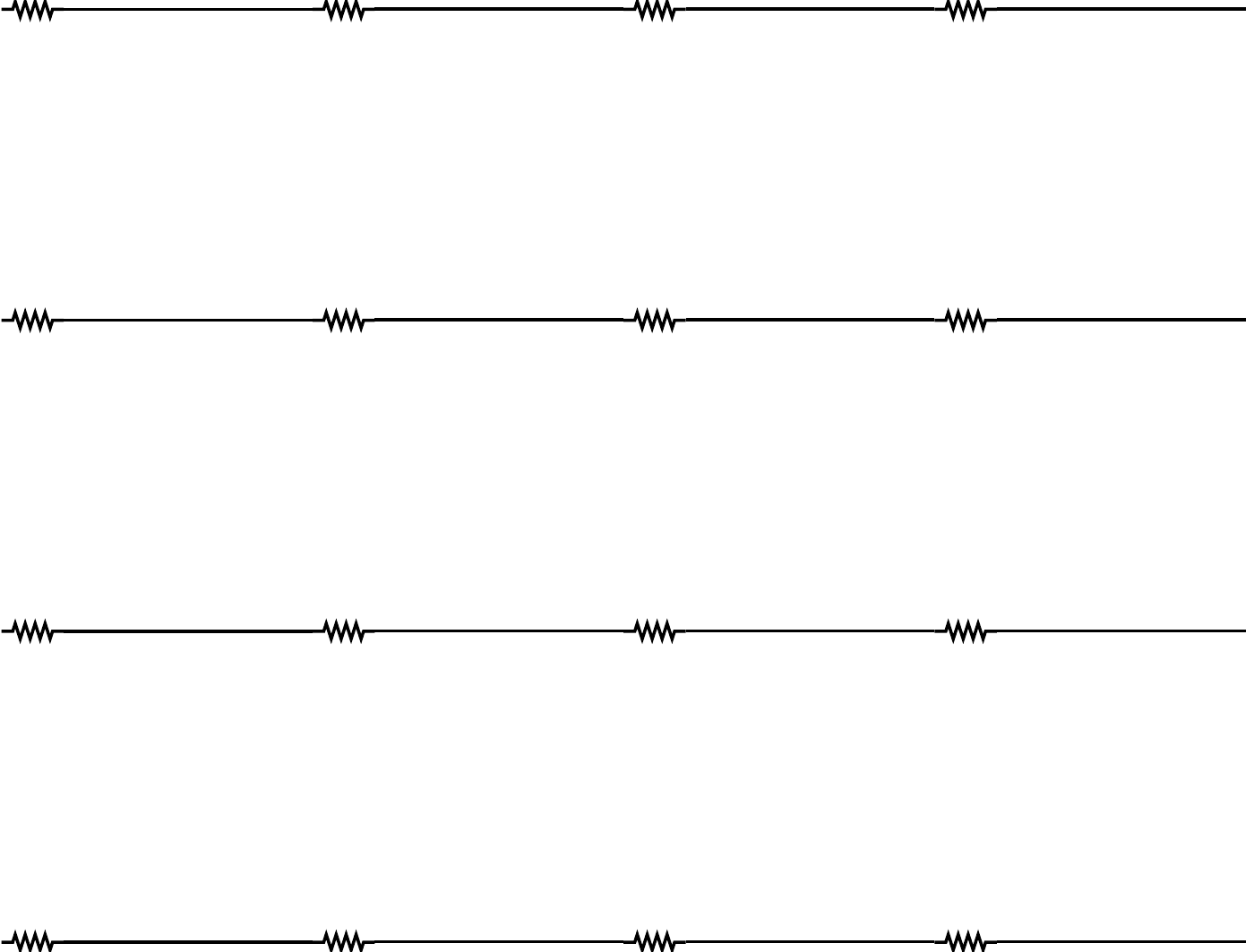}}
\caption{A division of MCA crossbar}
\label{fig:div}
\end{figure}

\begin{figure}[ht]
\centering
    \includegraphics[scale=0.45]{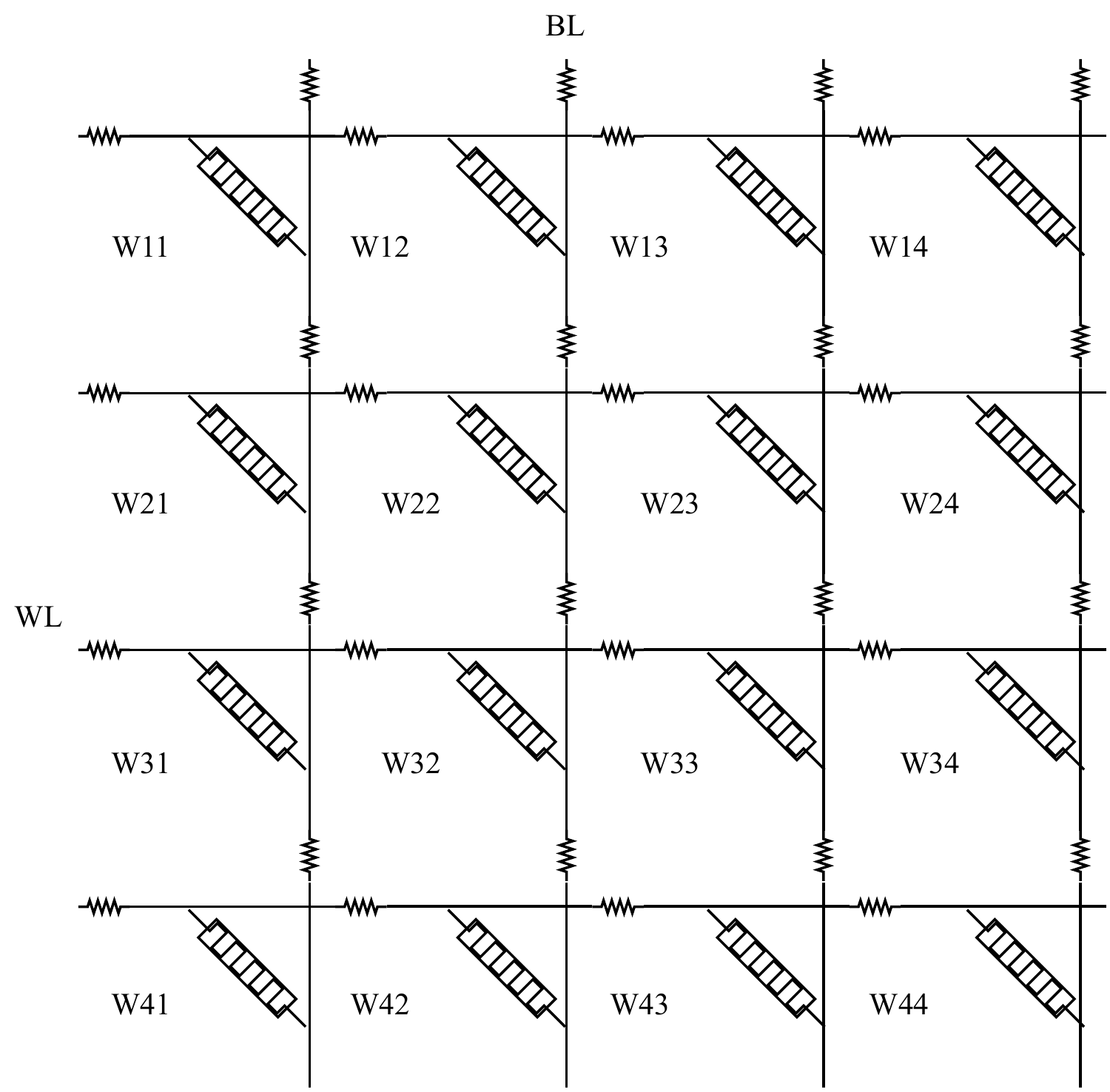}
\caption{A general MCA is shown with BL (bit line) and WL (word line).}
\label{fig:MCA}
\end{figure}

A voltage-controlled MCA is illustrated by Fig \ref{fig:MCA}. It can be divided into three parts: the top metal layer, the middle vertical memristor devices and the bottom metal layer, as shown in Fig.~\ref{fig:div_mem}, Fig.~\ref{fig:div_col} and Fig.~\ref{fig:div_row}. The top and the bottom metal layers are assumed to be two uniform grids, with equal conductance for each grid segment (but the conductance per segment can be different for the two layers). The memristor devices lie between the corresponding points of the two grids.

% The driven voltage sources are applied on the left boundary of the top layer, and all the other boundaries are terminated with a resistor. 

% In Fig \ref{fig:MCA}, analog input vector values are mapping to the voltage with alterable pulse lengths and amplitudes voltages, and the output vector values are represented as currents.

% The operation of MCA is divided into two parts, writing process and reading process. As there are no insulation between reading and writing, crosstalk is existed in the circuit, which we would not pay attention to in this article.

% \subsection{Steady-State MNA Formulation}
% In a SPICE model, the crossbar array circuits are commonly illustrated by matrix.
The steady-state MNA equation is given in \eqref{eq:MNA}, where $G_t$ and $G_b$ are the conductance matrices for the top and the bottom layers. $V_t$ and $V_b$ are the corresponding nodal voltage unknowns. $I_t$ and $I_b$ are the nonlinear functions of $V_t$ and $V_b$ relating the steady-state memristor currents to the applied voltages. Additionally, $Y_t$ and $Y_b$ are the boundary conditions. All of them combine to form the matrix equation \eqref{eq:F}. The whole nonlinear equation is solved by the Newton's method \eqref{eq:Newton3} with the Jacobian matrix given in \eqref{eq:J}.

\begin{subequations}
    \begin{equation}
    	G_t\ast V_t+I_t(V_t,V_b)-Y_t=0
    \end{equation}
    \begin{equation}
    	G_b\ast V_b-I_b(V_t,V_b)-Y_b=0
    \end{equation}
    \label{eq:MNA}
\end{subequations}
\begin{subequations}
\begin{equation}
    G_t=I\otimes G
    \label{eq:Gt}
\end{equation}
\begin{equation}
    G_b=G\otimes I
    \label{eq:Gb}
\end{equation}    
\end{subequations}
\begin{equation}
    \ J\left(\vec{V_n}\right)\ast\left(\vec{V_{n+1}}-\vec{V_n}\right)=\left(-F\left(\vec{V_n}\right)\right)
    \label{eq:Newton3}
\end{equation}
\begin{equation}
   J = F^\prime=\left[
    \begin{matrix}
    G_t&0\\[2mm]0&G_b\\
    \end{matrix}\right]+\left[
    \begin{matrix}\frac{\partial I_t}{\partial V_t}&\frac{\partial I_t}{\partial V_b}\\[2mm]\frac{-\partial I_b}{\partial V_t}&\frac{-\partial I_b}{\partial V_b}\\
    \end{matrix}
    \right]
    \label{eq:J}
\end{equation}

% We give our model by the following MNA (Modified Nodal Analysis) equation \eqref{eq:MNA}, where the $t$ and $b$ in subscript represented top layer and bottom layer of an MCA. $I_t$ and $I_b$ are function that related to voltage applied to the nodes from bottom and top wire, a set of nonlinear equations. $V_t$ is an array of the top nodal voltages arranged by rows, as well as $V_b$, shown in (3). The $v_{ti,j}$ and $v_{bi,j}$ represent the nodal voltage of the $i^{th}$ row and $j^{th}$ column on the top and bottom grid, respectively. The $G$ represent the wire conductance matrix in MNA (4). Additionally, (5a) and (5b) showcase the conductance matrices of top and bottom, where $G_t$ and $G_b$ are quasi-tri-diagonal matrix. $Y_t$ and $Y_b$ are the non-ideal properties in the circuit of top layer and bottom layer, usually the writing and reading noise and thermal noise.

% \subsection{Coupling Model}

% Since both layer’s nodal voltage in MCA will affect the memristor respectively, it is a non-linear relationship, so we couple the two equations. Combining the two equations into one matrix better for machine solving. A coupled matrix equation is proposed in (6).

% To approach the result of (7).
% Newton-Raphson method is applied, in every iteration, shown in (8).
% The Jacobian matrix of $F$ is $F'$, shown in (9).
% Hence the equation becomes to (10).
% Further get:
% (11).
% The equation of our primary concern has been transformed into
% (12).

\section{The Proposed Preconditioning Technique}

In this work we focus on using the iterative solution method of GMRES (Generalized minimal residual method) to solve the sparse total Jacobian matrix in \eqref{eq:J}. $J$ consists of two parts: the linear conductance matrix from the interconnect and the nonlinear Jacobian from the I-V functions of the memristor devices.   
\subsection{Preconditioner Formulation}
Firstly, we choose a particular indexing scheme to give $J$ a special sparsity structure. The top and the bottom layers both use natural indexing, but the directions are perpendicular to each other, as illustrated in Fig.~\ref{fig:div}. There are two reasons for this choice: 1) the four blocks in the nonlinear Jacobian matrix are now all diagonal; 2) by assuming equal conductances for all segments at the same layer, we can rewrite the top and the bottom linear conductance matrices $G_t, G_b \in \mathcal{R}^{n^2\times n^2}$ into Kronecker products~\eqref{eq:Gt} and \eqref{eq:Gb}, where $G\in \mathcal{R}^{n\times n}$ is the conductance matrix of single row or column~\eqref{eq:G}.

% Fig.~\ref{fig:A} show the sparsity pattern of $J$ for a MCA with $n = 8$ and with a sparse matrix dimension of $128\times128$.  
\begin{equation}
    P=\left[\begin{matrix}G_t+a_1I&-a_1I\\-a_2I&G_b+a_2I\\\end{matrix}\right]
    \label{eq:P}
\end{equation}

% In the process of matrix selection, to ensure that the basic characteristics of the original matrix (shape of the matrix inverse and range) are not lost, and at the same time to be able to perform the inversion well, we proposed a method based on 4-block inversion and matrix vector multiplication.

Next, we develop a special preconditioner of the form in~\eqref{eq:P} with the same block structure. The $a_1$ and $a_2$ are the mean of the diagonal elements of $\frac{\partial I_t}{\partial V_b}$ and $\frac{-\partial I_b}{\partial V_b}$, which can be considered as the average conductance of the memristor devices. Notice that $\frac{\partial I_t}{\partial V_t}$ and $\frac{\partial I_t}{\partial V_b}$ are opposite, as well as $\frac{-\partial I_b}{\partial V_t}$ and $\frac{-\partial I_b}{\partial V_b}$, since $V_b$ and $V_t$ are the voltages across the memristors.

\subsection{Fast Evaluation of Preconditioner}
\begin{equation}
% \begin{split}
\left[\begin{matrix}A&B\\C&D\\\end{matrix}\right]^{-1}=
\left[\begin{matrix}-M^{-1}DB^{-1}&M^{-1}\\B^{-1}+B^{-1}AM^{-1}DB^{-1}&-B^{-1}AM^{-1}\\\end{matrix}\right]
% \end{split}
\label{eq:block}
\end{equation}
\begin{equation}
\begin{aligned}
    M&=(C-DB^{-1}A)\\
    &=(-a_2I-(G_b+a_2I){(-a_1I)}^{-1}(G_t+a_1I))\\
    &=(-a_2I+\frac{1}{a_1}(G\otimes I+a_2I)(I\otimes G+a_1I))\\
    &=(-a_2I+\frac{1}{a_1}(G_2\otimes I)(I\otimes G_1))\\
    &=(-a_2I+\frac{1}{a_1}(G_2\otimes G_1))
\end{aligned}
\label{eq:M}
\end{equation}

It is important to have a fast scheme to evaluate $P^{-1} v$. We first apply the Woodbury block matrix inversion identity \eqref{eq:block}. Note that the off-diagonal blocks $B$ and $C$ are just identity matrices whose inverse is trivial. The core operation is to obtain $M^{-1}v=(C-DB^{-1}A)^{-1}v$. 

To this end, we rewrite $M$ into \eqref{eq:M}, with $G_1$ and $G_2$ given in \eqref{eq:G1} and \eqref{eq:G2}. In typical MCAs, the memristor conductance is generally much smaller than that of interconnects. Therefore, one can drop the first term on the right hand side of \eqref{eq:diagM} and approximate $M$ as in \eqref{eq:M-1hat} and \eqref{eq:G2hat}. To compute \eqref{eq:2matrix1}, where ${\hat{g}}_{i,j}^2$ is the element of $\widehat{G_2}$. Vector $v$ can be rearranged by \eqref{eq:2matrix2}, $\widehat{V_j}$ represent the $j^{th}$ column of $\hat{V}$. Consider the $j^{th}$ row in \eqref{eq:2matrix3}.
Finally, we can deduce original equation to \eqref{eq:2matrix4}.

% In the simplified equation we consider only the elements on the diagonal, and the diagonal elements of M are all equal, as \eqref{eq:diagM}.
% Here we take an approximation of $M$ for easy inversion, omitting the small value $a_2$, shown in \eqref{eq:condition}.
% Therefore, it comes to \eqref{eq:G2hat}. Due to the feature of tensor product, we can get $M^{-1}$ in \eqref{eq:M-1hat}.

% \subsection{Transformation of Matrix Multiplication}

% The preconditioner in the GMRES algorithm is passed into each iteration in the form of matrix-vector product, so we can calculate the matrix and vector multiplication from right to left to simplify the calculation steps.

% This conversion of linear equations into matrix equations can effectively avoid the calculation of large-dimensional matrices derived from tensor products.

\section{Numerical Results}
In the following tests, the top and bottom wire conductance $g$ per segment are normalized to $1$. We adopt the Yakopcic model~\cite{Yakopcic:11} as the RRAM model. Since the proposed method is expected to handle RRAM devices of various states, we obtain the conductance matrix of RRAM by randomly setting the internal state variable of their model, with a maximum conductance being $0.4$ to meet the approximation condition~\eqref{eq:condition}. The GMRES solver from Scipy is used with a uniform relative tolerance of ${10}^{-6}$. 
% All the tests are done on a server with Intel® Core™ i7-10700 CPU @ $2.9$GHz with $24$Gb Memory.

% It should be noted that in the left preconditioning GMRES algorithm, $\xi$ in the stopping criterion is not a modulus of the true residuals of the original system of equations, but a modulus of the preprocessed residuals $r^k\ =\ P^{-1}(b\ -\ Ax^{(k)})$. There is no better way to compute the true residuals than to compute them explicitly. Therefore, the residuals calculated here are not the true residuals.

\begin{figure}[ht]
    \centering
    \includegraphics[width=0.45\textwidth]{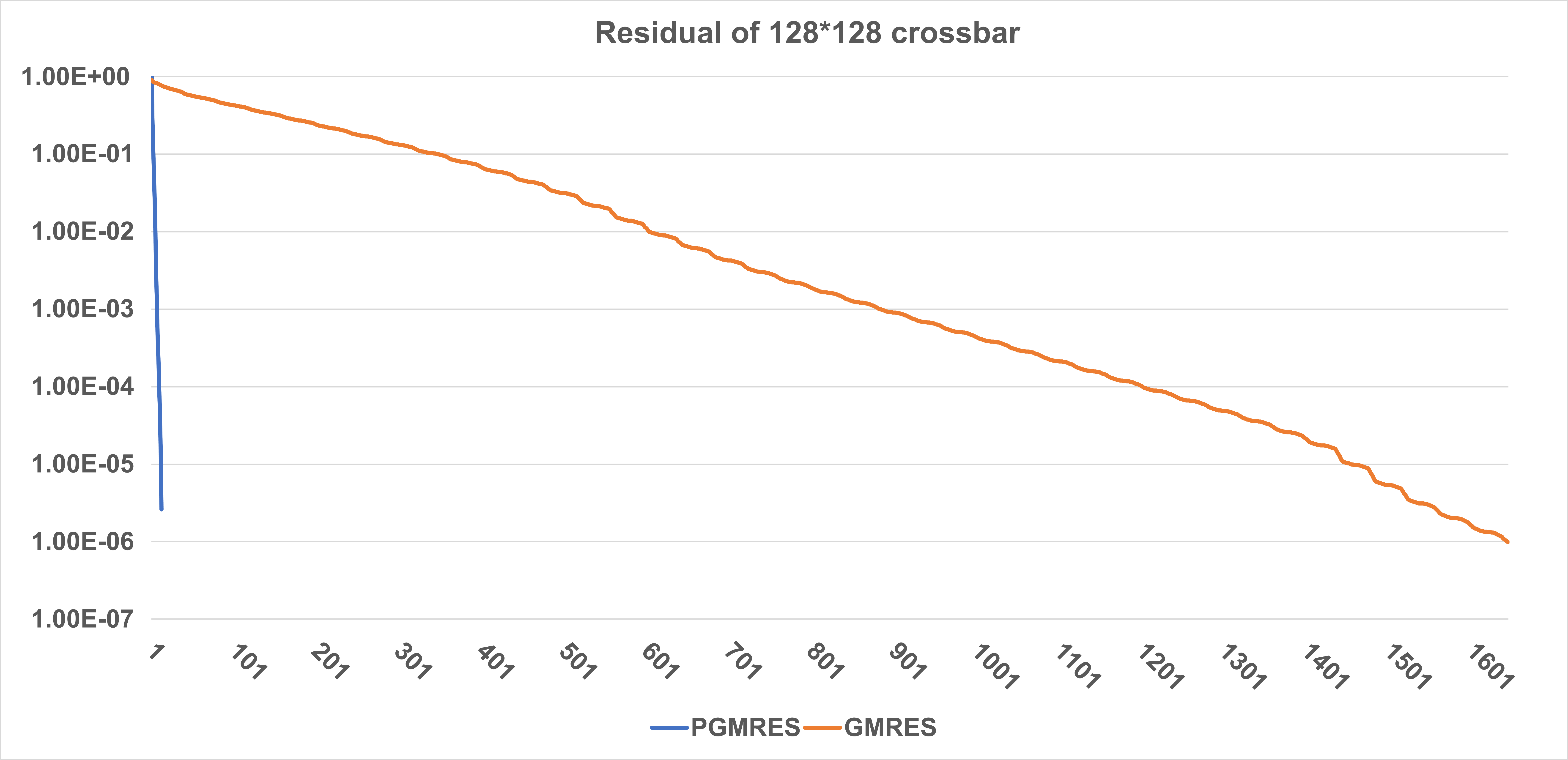}
    \caption{Residual of preconditioned GMRES (PGMRES) and baseline GMRES for $128\times128$ crossbar .}
    \label{fig:res128}
\end{figure}

Fig.~\ref{fig:res128} shows the residual history of GMRES with and without the proposed preconditioner. The test case is a $128\times128$ crossbar with the matrix dimension of $32768\times32768$. It can be seen that the proposed preconditioner drastically accelerates the convergence of GMRES. 

\begin{figure}[ht]
    \centering
    \includegraphics[width=0.45\textwidth]{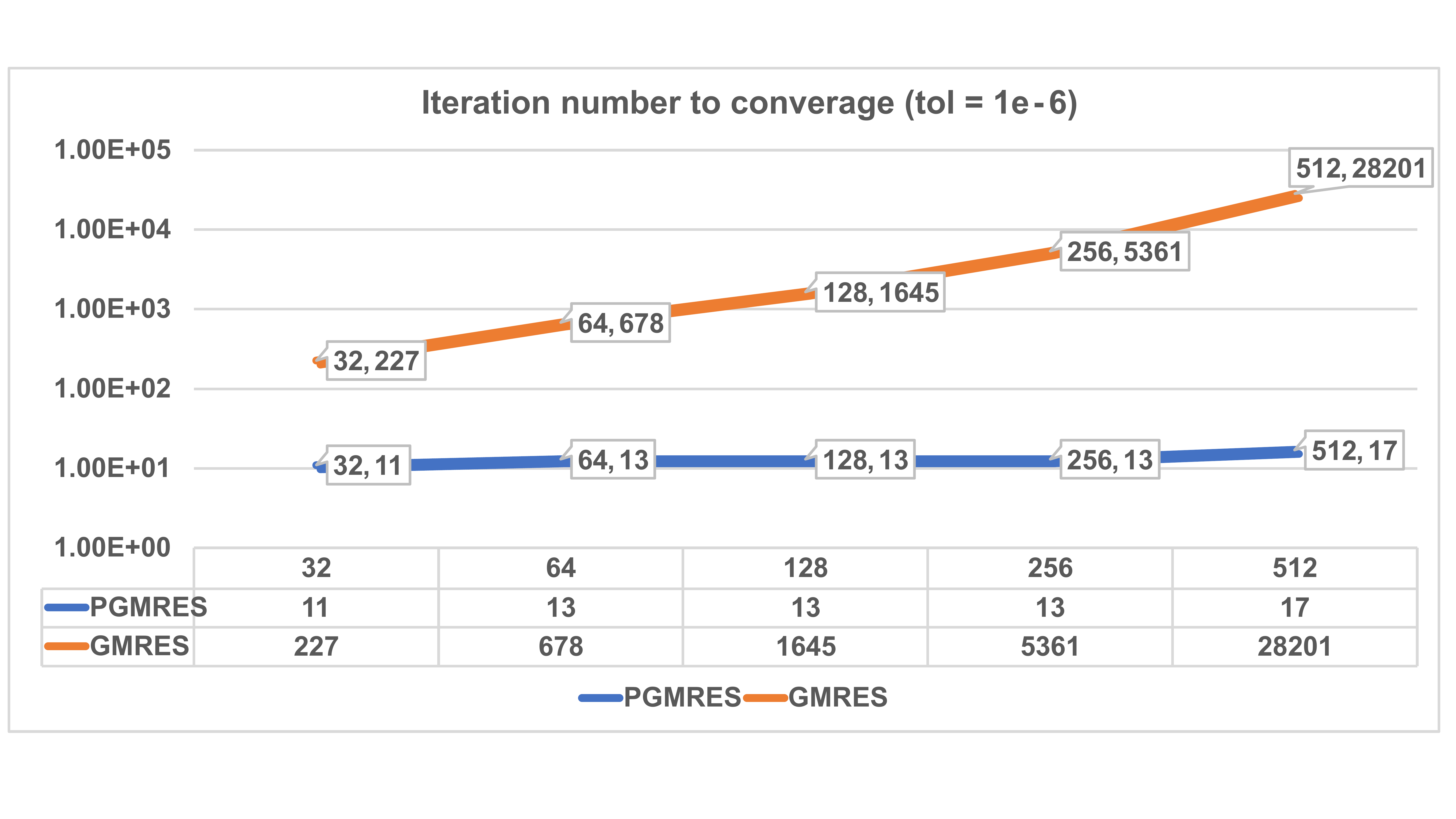}
    \caption{Iteration number to coverage of dimension of crossbar $32\times32$, $64\times64$, $128\times128$, $256\times256$ and $512\times512$.}
    \label{fig:iternum}
\end{figure}

Fig.~\ref{fig:iternum} compares the iteration number for MCAs of five difference sizes ($32\times32$, $64\times64$, $128\times128$, $256\times256$ and $512\times512$). The matrix sizes are labeled on the lines and the corresponding iteration numbers summarized in the table. It is clear that the computational saving from the proposed preconditioner grows rapidly as the matrix size increases.   

% \begin{table}[h]
% 	\caption{Comparison of Total CPU Time Consumption and Iteration Steps to Coverage}
% 	\centering
% \resizebox{0.5\textwidth}{!}{%
% \begin{tabular}{ccccc}
% \hline
%                       & \multicolumn{2}{c}{After Preconditioned} & \multicolumn{2}{c}{Before Preconditioned} \\ \hline
% Crossbar Dimension (n) & Steps      & CPU time consumption/s      & Steps       & CPU time consumption/s      \\ \hline
% 16*16                  & 11         & 0.00299                     & 67          & 0.01596                     \\ \hline
% 32*32                  & 11         & 0.02194                     & 227         & 0.05785                     \\ \hline
% 64*64                  & 13         & 0.09275                     & 678         & 0.30377                     \\ \hline
% 128*128                & 13         & 0.57907                     & 1645        & 1.97858                     \\ \hline
% 256*256                & 13         & 4.77912                     & 5361        & 13.41768                    \\ \hline
% 512*512                & 17         & 225.12859                   & 22801       & 331.31852                   \\ \hline
% \end{tabular}%
% }
% \label{table:comparison}
% \end{table}

% Please add the following required packages to your document preamble:
% \usepackage{graphicx}
\begin{table}[ht]
\caption{Comparison of Total CPU Time Consumption and Iteration Steps to Coverage}
\centering
% \large
\resizebox{.5\textwidth}{!}{% 
\setlength{\tabcolsep}{.5mm}
\begin{tabular}{ccccccccc}
\toprule
 &
  \multicolumn{2}{c}{\textbf{\begin{tabular}[c]{@{}c@{}}Before\\ Preconditioned\end{tabular}}} &
  \multicolumn{2}{c}{\textbf{\begin{tabular}[c]{@{}c@{}}Jacobi\\ Preconditioner\end{tabular}}} &
  \multicolumn{2}{c}{\textbf{\begin{tabular}[c]{@{}c@{}}ILU\\ Preconditioner\end{tabular}}} &
  \multicolumn{2}{c}{\textbf{\begin{tabular}[c]{@{}c@{}}Our\\ Preconditioner\end{tabular}}} \\ \midrule
\begin{tabular}[c]{@{}c@{}}Crossbar\\ Dimension (n)\end{tabular} &
  Steps &
  \begin{tabular}[c]{@{}c@{}}CPU time\\ consumption/s\end{tabular} &
  Steps &
  \begin{tabular}[c]{@{}c@{}}CPU time\\ consumption/s\end{tabular} &
  Steps &
  \begin{tabular}[c]{@{}c@{}}CPU time\\ consumption/s\end{tabular} &
  Steps &
  \begin{tabular}[c]{@{}c@{}}CPU time\\ consumption/s\end{tabular} \\ \midrule
16*16   & 67    & 0.01596   & 60    & 0.00897   & 3    & 0.00598   & 11 & 0.00299   \\ \midrule
32*32   & 227   & 0.05785   & 203   & 0.03092   & 15   & 0.01396   & 11 & 0.02194   \\ \midrule
64*64   & 678   & 0.30377   & 551   & 0.14319   & 111  & 0.14561   & 13 & 0.09275   \\ \midrule
128*128 & 1645  & 1.97858   & 1610  & 0.95511   & 308  & 1.24064   & 13 & 0.57907   \\ \midrule
256*256 & 7526  & 28.51413  & 5361  & 23.41768  & 589  & 11.05555  & 13 & 4.77912   \\ \midrule
512*512 & 22801 & 331.31852 & 19273 & 324.20583 & 3120 & 309.91122 & 17 & 225.12859 \\ \bottomrule
\end{tabular}%
}
\label{table:comparison}
\end{table}
Table~\ref{table:comparison} compares the proposed preconditioner against other mainstream preconditioners such as the Jacobi and the ILU preconditioner. The iteration number and the total CPU runtime are recorded for MCAs of different sizes. For small cases, the three types of preconditioners perform comparably well. For larger cases, the proposed preconditioner requires much fewer iterations than the other two preconditioners. The runtime reduction is less significant due to the evaluation of preconditioner not being fully optimized. Future efforts will be devoted to speed up this part. 

\section{Conclusion}
We have devised an efficient preconditioner for fast iterative solution of the Jacobian matrices appearing in steady-state MCA simulation. The preconditioner leverages the special sparsity pattern in the Jacobian matrices resulted from a deliberately crafted indexing scheme. Tensor product and block matrix inversion techniques are utilized to significantly accelerate the preconditioner evaluations during the iterative solutions. Numerical results have demonstrated the efficacy of the proposed preconditioner.

\ifCLASSOPTIONcaptionsoff
  \newpage
\fi

% 在thebibliography中按格式插入参考文献

\clearpage
\subsection{Equations and Figures}
\small
% \begin{subequations}
%     \begin{equation}
%     	i=G(x,v)v 
%     \end{equation}
%     \begin{equation}
%     	G(x,0)\neq\infty
%     \end{equation}
%     \begin{equation}
%     	\frac{dx}{dt}=g(x,v)
%     \end{equation}
%     \end{subequations}
    
\begin{equation}
    G=\left[\begin{matrix}\begin{matrix}2g&-g\\-g&2g\\\end{matrix}&\cdots&0\\\vdots&\ddots&\vdots\\0&\cdots&\begin{matrix}2g&-g\\-g&2g\\\end{matrix}\\\end{matrix}\right]_{n\ast n}
    \label{eq:G}
\end{equation}

% \begin{equation}
% V_t=\left[\begin{matrix}v_{t1,1}\\v_{t1,2}\\\vdots\\v_{t1,n-1}\\v_{t1,n}\\v_{t2,1}\\\vdots\\v_{t2,n}\\\vdots\\v_{tn,n}\\\end{matrix}\right],\ V_b=\left[\begin{matrix}v_{b1,1}\\v_{b1,2}\\\vdots\\v_{b1,n-1}\\v_{b1,n}\\v_{b2,1}\\\vdots\\v_{b2,n}\\\vdots\\v_{bn,n}\\\end{matrix}\right]
% \label{eq:V}
% \end{equation}

\begin{equation}
    F=\ \left[\begin{matrix}G_t&0\\0&G_b\\\end{matrix}\right]*\ \left[\begin{matrix}V_t\\V_b\\\end{matrix}\right]+ \left[\begin{matrix}I_t\\{-I}_b\\\end{matrix}\right]- \left[\begin{matrix}Y_t\\Y_b\\\end{matrix}\right] = 0
    \label{eq:F}
\end{equation}

% \begin{equation}
%     F=0
%     \label{eq:F=0}
% \end{equation}

% \begin{equation}
%     \vec{V_{n+1}}=\vec{V_n}- F\left(\vec{V}\right)/J\left(\vec{V}\right)
%     \label{eq:Newton1}
% \end{equation}

% \begin{equation}
%     \vec{V_{n+1}}-\vec{V_n}=\ J^{-1}\left(\vec{V_n}\right)\ast\left(-F\left(\vec{V_n}\right)\right)
%     \label{eq:Newton2}
% \end{equation}

% \begin{equation}
%     Ax=b.
%     \label{eq:Ax=b}
% \end{equation}

% \begin{equation}
%     P^{-1}Ax=P^{-1}b
%     \label{eq:Ax=b,precondition}
% \end{equation}

\begin{subequations}
    \begin{equation}
        G_1=\left[\begin{matrix}\begin{matrix}2g+a_1&-g\\-g&2g+a_1\\\end{matrix}&\cdots&0\\\vdots&\ddots&\vdots\\0&\cdots&\begin{matrix}2g+a_1&-g\\-g&2g+a_1\\\end{matrix}\\\end{matrix}\right]_{n\ast n}
        \label{eq:G1}
    \end{equation}
    \begin{equation}
        G_2=\left[\begin{matrix}\begin{matrix}2g+a_2&-g\\-g&2g+a_2\\\end{matrix}&\cdots&0\\\vdots&\ddots&\vdots\\0&\cdots&\begin{matrix}2g+a_2&-g\\-g&2g+a_2\\\end{matrix}\\\end{matrix}\right]_{n\ast n}
        \label{eq:G2}
    \end{equation}
\end{subequations}

\begin{equation}
    diag(M)=(-a_2+\frac{1}{a_1}(2g+a_2)(2g+a_1))
    \label{eq:diagM}
\end{equation}

\begin{equation}
    20a_2<\frac{1}{a_1}(2g+a_2)(2g+a_1)
    \label{eq:condition}
\end{equation}

\begin{equation}
    M\cong\hat{M}=\widehat{G_2}\otimes G_1
    \label{eq:Mhat}
\end{equation}

\begin{equation}
    \widehat{G_2}=\left[\begin{matrix}\begin{matrix}\frac{2g+a_2}{a_1}&-g\\-g&\frac{2g+a_2}{a_1}\\\end{matrix}&\cdots&0\\\vdots&\ddots&\vdots\\0&\cdots&\begin{matrix}\frac{2g+a_2}{a_1}&-g\\-g&\frac{2g+a_2}{a_1}\\\end{matrix}\\\end{matrix}\right]_{n\ast n}
    \label{eq:G2hat}
\end{equation}

\begin{equation}
    M^{-1}\cong{\hat{M}}^{-1}={\widehat{G_2}}^{-1}\otimes{G_1}^{-1}
    \label{eq:M-1hat}
\end{equation}

\begin{equation}
\begin{aligned}
    &\left({\widehat{G_2}}^{-1}\otimes{G_1}^{-1}\right)v\\
    &=\left({\widehat{G_2}}^{-1}\otimes{G_1}^{-1}\right)\left[\begin{matrix}v_{1,1}\\v_{1,2}\\\vdots\\v_{1,n-1}\\v_{1,n}\\v_{2,1}\\\vdots\\v_{2,n}\\\vdots\\v_{n,n}\\\end{matrix}\right]\\
    &=\left[\begin{matrix}\frac{1}{{\hat{g}}_{1,1}^2}\ast{G_1}^{-1}&\cdots&\frac{1}{{\hat{g}}_{1,n}^2}\ast{G_1}^{-1}\\\vdots&\ddots&\vdots\\\frac{1}{{\hat{g}}_{n,1}^2}\ast{G_1}^{-1}&\cdots&\frac{1}{{\hat{g}}_{n,n}^2}\ast{G_1}^{-1}\\\end{matrix}\right]\left[\begin{matrix}v_{1,1}\\v_{1,2}\\\vdots\\v_{1,n-1}\\v_{1,n}\\v_{2,1}\\\vdots\\v_{2,n}\\\vdots\\v_{n,n}\\\end{matrix}\right]
\end{aligned}
\label{eq:2matrix1}
\end{equation}

\begin{equation}
    \hat{V}=\left[\begin{matrix}v_{1,1}&\cdots&v_{1,n}\\\vdots&\ddots&\vdots\\v_{n,n}&\cdots&v_{n,n}\\\end{matrix}\right]_{n\ast n}=\left[\begin{matrix}\widehat{V_1}&\cdots&\widehat{V_n}\\\end{matrix}\right]
    \label{eq:2matrix2}
\end{equation}
\begin{equation}
    \begin{aligned}
        &\frac{1}{{\hat{g}}_{j,1}^2}\ast{G_1}^{-1}\widehat{V_1}+\frac{1}{{\hat{g}}_{j,2}^2}\ast{G_1}^{-1}\widehat{V_2}+\ldots+\frac{1}{{\hat{g}}_{j,n}^2}\ast{G_1}^{-1}\widehat{V_n}\\
      & =\left[\begin{matrix}{G_1}^{-1}\widehat{V_1}&\ldots&{G_1}^{-1}\widehat{V_n}\\\end{matrix}\right]\left[\begin{matrix}\frac{1}{{\hat{g}}_{j,1}^2}\\\vdots\\\frac{1}{{\hat{g}}_{j,n}^2}\\\end{matrix}\right]
    \end{aligned}
    \label{eq:2matrix3}
\end{equation}
\begin{equation}
    \left({\widehat{G_2}}^{-1}\otimes{G_1}^{-1}\right)v={G_1}^{-1}\hat{V}{G_2}^{-1}
    \label{eq:2matrix4}
\end{equation}

% \subsection{Figures}
%

% \begin{figure}[ht]
%     \centering
%     \includegraphics[width=0.4\textwidth]{fig/fig3.pdf}
%     \caption{Schematic diagram of the shape of matrix A when $n=8$. Note that dimension of the matrix here is $\left(2\ast n^2\right)\times\left(2\ast n^2\right)$.}
%     \label{fig:A}
% \end{figure}

\end{document}